# Thick collagen-based matrices including growth factors to induce 3D neural differentiation

*Marie-Noëlle Labour[1,2,3], Amélie Banc[1,2], Audrey Tourette[1,2], Frédérique Cunin[2], Jean-Michel Verdier[1,3], Jean-Marie Devoisselle[2], Anne Marcilhac[1,3] and Emmanuel Belamie[*1,2].*

[1] Ecole Pratique des Hautes Etudes, 46 rue de Lille, 75007 Paris

[2] Institut Charles Gerhardt, UMR 5253, Equipe MACS, Ecole Nationale Supérieure de Chimie, 8 rue de l'école normale, F-34 296 Montpellier, Cedex 5;

[3] INSERM U 710 - Unité mixte UM2 - EPHE, Université Montpellier II - Place Eugène Bataillon-CC105, 34095 Montpellier Cedex 05

\* Corresponding author

Phone: +33 467 163 444, Fax: +33 467 163 470

E-mail address: emmanuel.belamie@enscm.fr

**Abstract**

Designing synthetic microenvironments for cellular investigations is a very active area of research at the crossroad of cell biology and materials science. The present work describes the design and functionalization of a 3D culture support dedicated to the study of neural cells differentiation. It is based on a dense self-assembled collagen matrix stabilized by 100 nm-wide interconnected native fibrils, without chemical cross-linking. The matrices were made suited for cell manipulations and direct observation in confocal microscopy by anchoring them to traditional glass supports, with a calibrated thickness of *c.a.* 50 µm. The matrix composition can be readily adapted to specific neural cell types, notably by incorporating appropriate neurotrophic growth factors. PC-12 and SH-SY5Y lines both exhibit high differentiation rates (up to 66% and 49% respectively) with the corresponding growth factors (500 ng NGF, 50 ng BDNF) simply impregnated and slowly released from the support. We also show that both growth factors can be chemically conjugated (EDC/NHS) throughout the matrix and yield similar differentiation rates (61% and 52% respectively). Finally, we observed neurite outgrowth over several tens of microns within the 3D matrix, with both diffusing and immobilized growth factors, which raises the possibility of designing simplified brain tissue model materials.

## 1-Introduction

Cellular microenvironments, including the structure and rigidity of the substrate, topological and chemical cues like growth factor concentration, orient the cell faith and behavior in physiological conditions. Smart devices for *in vitro* investigations of cellular events involved in their functions and in pathological situations must include a set of these stimuli associated to a spatio-temporal control of their application to the cells [1, 2]. The response to such cues gives an opportunity to orient the behavior of cells, favor a particular phenotypic engagement, and study specific aspects of their functions and dysfunctions. More specifically, a very active area of research is dedicated to the study and control of neural cell differentiation, and in particular the outgrowth and guidance of neurite, and the effects of distal events on the whole cell faith. Typical examples are the retrograde signaling triggered by ligand-receptor interactions on distant neurites [3, 4], or A$\beta$-induced neurite degeneration followed by cell death [5]. Systems developed to this purpose most often rely on microfabrication techniques to design patterns with adhesion molecules or growth factor gradients [6-12]. Such patterned substrates are useful to study the formation and branching of neurites, but also axonal dystrophies notably involved in neurodegenerative pathologies like Alzheimer disease (AD). Several compartmented systems have also been developed with the aim of studying the response of neurites separately from that of the rest of the cell [13-20]. However, in most cases, the cells mostly evolve in a 2D environment and/or the set-up does not permit the vizualization in 3D of cell protrusions like neuritic extensions.

Here, we have developed a 3D biomimetic compartmented system suited for the study of cellular events occurring within the matrix, near the tip of neurites. Macromolecules of the extracellular matrix (ECM) like collagen, fibronectin or laminin are choice substrates to enhance cell adhesion, and in some cases are required for the proper differentiation of specific cell types [21, 22]. Most often these adhesion molecules are deposited as a very thin layer onto the surface of glass or plastic dishes. The main drawbacks of such very thin coatings are that *i)* the mechanical properties sensed by the cells are those of the underlying material (glass, plastic), which can largely modify the cells behavior [23] and *ii)* the study of cellular events is mostly limited to two dimensions. Conversely, cells seeded on top of thick hydrogels comprised of either polysaccharides or proteins like collagen [24-27] are put in the presence of a 3D environment. Such configuration presents a strong potential for the study of long-range cell migration mechanisms, for instance in the context of metastasis formation or wound healing. However, these thick matrices require large amounts of purified ECM macromolecules and are often not amenable to *in situ* observations like confocal microscopy. Inclusion of cells within 3D macromolecular hydrogels (collagen, alginate, matrigel…) [28-31] is also a powerful approach to investigate specific cellular behavior like the ability to contract support matrices, the response to stress, the migration potential, etc. However, this well-proven technique has drawbacks, among which the stress generated by the encapsulation of cells during formation of the gel, and the fact that all components composing the gel are immediately in direct contact with the cells. Two-dimensional systems providing the cells with an adaptable three-dimensional network of physical and biochemical cues mimicking their natural niche [32] are a promising alternative to existing cell culture devices.

Here, we propose a system combining advantages of 3D culture supports, prepared by self-assembly with limited amounts of ECM macromolecules and an adaptable composition in relation with specific cell types. We improved the usability of collagen matrices as cell culture environments, suited for direct observations by confocal microscopy, by controlling their thickness and structure, and by chemically grafting them onto traditional glass supports. We show that different neurotrophic growth factors (Nerve growth factor, NGF and Brain derived neurotrophic factor, BDNF) can be included in the gels – impregnated or chemically conjugated – and effectively stimulate the differentiation of specific neural cell lines. PC-12 [33] and SH-SY5Y [34] lines both exhibit high differentiation rates, dependent upon the substrate and the actual amount of the corresponding growth factor accessible to the cells, released or covalently coupled to the gel. The possibility to elicit neurite outgrowth in a 3D context with diffusing or immobilized growth factor raises the possibility of designing simplified brain tissue model materials.

## 2. Materials and methods

### 2.1 Culture set-up preparation and optimization

*- Collagen extraction and purification*

Type I collagen was extracted and purified from young Wistar rat tails following a protocol previously described [35]. Briefly, tendons were excised and washed with phosphate-buffered saline (PBS) to remove cells and traces of blood. Tissues were then soaked in a 4 M NaCl solution, rinsed and solubilized in 500 mM acetic acid to lyse remaining cells and precipitate some of the high-molecular-weight proteins. The solution was clarified by filtration and centrifugation (21,000×g for 2 hours). Collagen was then precipitated in 600 mM NaCl, centrifuged (6000×g for 45 min at 10°C) and finally solubilized in 500 mM acetic acid. The solutions were dialyzed (1:10 vol.) at least twice against 500 mM acetic acid to equilibrate the solvent and centrifuged (21,000×g for 4h at 15°C) to remove remaining impurities or aggregate. A collagen solution at 45 mg/ml was prepared by reverse dialysis against 75 g/l polyethylene glycol (20-35 kDa, Fluka) dissolved in 500 mM acetic acid. Collagen concentration was determined from the acidic solution before fibrillogenesis by assessing the amount of hydroxyproline [36].

*- Grafting onto glass coverslips*

Covalent immobilization of collagen molecules on the glass surface of coverslips was carried out by successive surface modifications with an aminosysilane and a dialdehyde (Figure 1). Round coverslips (18 mm diameter) were thoroughly washed with methanol and distilled water before drying overnight at 110°C. They were then immersed in 2% Aminopropyltriethoxysilane (APTES from Sigma chemicals) in acetone for 3 hours at room temperature, and rinsed in acetone. Next, the coverslips were heated at 150°C for 1 hour to promote silanization of the surface. Then, the silanized surfaces were immersed in 5% glutaraldehyde (GA from Sigma Chemicals) in PBS for 2 hours at room temperature followed by rinsing with water, and drying in sterile conditions. Activated surfaces were later covered overnight at room temperature with an excess of collagen solution at 0.5 mg/ml (diluted with PBS from a 5 mg/ml collagen acetic acid solution). Grafted surfaces were washed with acetic acid (0.5 M), to remove the excess of ungrafted collagen molecules followed by deionized (milli-Q) water. Finally, reduction of Schiff's bases was achieved by immersing the coverslips into 0.02 M sodium borohydride (Sigma) in PBS for 1 hour at room temperature, followed by rinsing with deionized water.

- *Calibrated hydrogel scaffold formation*

A specially designed teflon mould was used to cast 50 μm-thick layers of acidic collagen solution (*c.a.* 45 mg/mL) onto collagen-grafted glass coverslips (Figure 1). A teflon lid was fixed and the whole setup submitted to ammonia vapours in a closed vessel to induce collagen fibrillogenesis. After three days, the samples were rinsed by immersing the mould for 12 hours in 0.25X PBS and freeze-dried. The samples were stored in a dried atmosphere and all manipulations were done in sterile conditions.

- *Growth factor impregnation and release*

Lyophilised dense collagen matrices were hydrated overnight in a wet atmosphere with 50 μl of Dulbecco's Modified Eagle Medium (DMEM, Gibco®) containing 500 ng of NGF (Nerve Growth Factor 2.5S, Murine, Promega). The excess of liquid was then removed and the scaffolds were rinsed in PBS. Growth factor release kinetics were determined by immersing the gels in 1 ml of DMEM supplemented with 0.1% BSA and 0.02% sodium azide, in 12-well plates under orbital shaking (200 rpm) at room temperature. At exponentially spaced time points, small volumes (10 or 15 μL) of supernatant were collected in triplicate and replaced by fresh medium. The amount of NGF released was quantified by a sandwich ELISA (NGF $E_{max}^{®}$ ImmunoAssay System, Promega). Briefly, ELISA plates (Nunc-Immuno$^{TM}$ plates MaxiSorp$^{TM}$, Fisher scientific) were coated with anti-NGF polyclonal antibody and blocked with the buffer provided. Soluble NGF from the sample is captured by the polyclonal antibody and further recognised by a monoclonal antibody. Finally, an HRP conjugated anti-immunoglobulin was added to the wells and after rinsing, the plates were incubated with TMB substrate and colour changes measured at 450 nm.

BDNF impregnation and release were performed in the same way. The medium was DMEM-F12 supplemented with 3% of Fetal Bovine Serum. BDNF (500 ng) was impregnated overnight at 4°C and the release was monitored at ambient temperature. Collected samples were kept at -20°C before quantification by sandwich ELISA (BDNF $E_{max}^{®}$ ImmunoAssay System, Promega).

- *Growth factor coupling*

Coupling of growth factors (GF), NGF and BDNF (recombinant human, Invitrogen) was performed on lyophilised gels first rehydrated with sterile PBS at pH 6. Activation was achieved by incubating the gels for approximately 8 hours at ambient temperature in a 33 mM EDC (1-Ethyl-3-(3-dimethylaminopropyl)carbodiimide, Sigma) and 66 mM NHS (N-hydroxysuccinimide, Sigma) solution in PBS at pH 6. After rinsing at least three times (10 min. in 1 ml of PBS, pH 7), collagen gels were kept immersed overnight at ambient temperature under orbital shaking at 200 rpm, in 1 ml of PBS at pH 7 containing 50 ng, 500 ng or no (control) growth factor. The medium was eliminated and the GF-coupled collagen gels were rinsed four times for at least 4 hours with 1 ml PBS pH 7 before plating the cells.

For colocalization experiments to evaluate the efficiency of the EDC/NHS coupling strategy, the coupling reaction was performed with collagen in solution. One ml of a 2 mg/ml collagen solution in MES buffer (pH 4.5) was mixed with 1 ml of 0.1M EDC-0.2M NHS solution in MES buffer and incubated with shaking for 30 minutes. The reaction was stopped with 0.4 M 2-mercaptoethanol for 15 minutes under shaking before adding NGF solution (shaking overnight at ambient temperature). Dialysis (Cellu•Sep® membrane, cut-off 6 kDa) was performed against PBS pH 7.4 for 4 hours to remove the remaining reagents and adjust the mixture to neutral pH to complete the reaction with NGF. Non-conjugated NGF was then removed by a second dialysis (Spectra/Por® Biotech membranes, cut-off: 50 kDa) and two successive centrifugations (30 min., 3,000×g). The pellet containing NGF-conjugated collagen fibres was placed on a microscopy slide and incubated with an anti-NGF monoclonal antibody (Promega) followed by an AlexaFluor® 594 (Invitrogen) secondary antibody for detection. Fluorescence and Differential Interference Contrast pictures were acquired on a Zeiss Axioimager microscope at the Montpellier RIO Imaging Facility.

**2.2 Materials characterization.**

*- Polarization Modulation - InfraRed Reflexion Absorption Spectroscopy (PM-IRRAS)*
Experiments were carried out on 1 millimeter-thick grafted glass slides, in order to avoid interferences due to the low thickness of coverslips. PM-IRRAS spectra were recorded on a Nicolet 870 Fourier transform infrared (FT-IR) spectrometer by co-addition of 600 scans at a resolution of 8 cm$^{-1}$. The details of the optical setup, the experimental procedure, and the two-channel processing of the detected intensity have been described elsewhere [37, 38]. All the spectra presented in this paper result from the ratio of the sample spectra to the substrate spectrum.

*- Contact angle*

Water contact angle measurements were performed using a Digidrop GBX fast/60 apparatus.

*- Breaking strength*

The efficiency of collagen grafting was estimated by performing ultimate tensile strength measurements on dense collagen matrices sandwiched between two grafted coverslips with a texture analyser apparatus TA.XT2 (Stable Micro Systems, Ltd in Godalming, Surrey UK) operated in the traction mode. The collagen solution (45 mg/ml in acetic acid) was deposited with a thickness of about 1 mm defined by glass spacers between two grafted coverslips and was submitted to ammonia vapours overnight to induce fibrillogenesis. The samples were attached to the static and moving parts of the apparatus, by means of metal connectors glued to the free glass surfaces on each side. Tensile tests were performed with a constant velocity of 0.1 mm/s, and the ultimate breaking strength was determined.

*- Scanning Electron Microscopy (SEM)*

For SEM observations, collagen scaffolds were previously fixed by immersion in a glutaraldehyde solution (2.5% in PBS) for 1 hour at ambient temperature and dehydrated in water/ethanol mixtures (10%, 30%, 50%, 70% and 100% ethanol). Samples were then dried by supercritical drying with $CO_2$ (E3100 Polaron Critical Point drier). After platinum-sputter coating (Balzers SCD 020) samples were observed in the secondary electron mode with a scanning electron microscope (Hitachi 4800 S) operated at an accelerating voltage of 50 kV. In order to prevent gel collapse and measure gels thickness, edges of samples were observed under low vacuum (0.7 torr) with a Quanta 200F FEI microscope at an accelerating voltage of 15 kV.

*- Growth factor distribution through the gel*

To characterize the spatial distribution of the coupled growth factor (NGF), gels were fixed at room temperature 1 hour with 4% paraformaldehyde (PFA) and incubated in 30% sucrose overnight. The samples were then placed face down on moulds filled with Tissue-Tek® O.C.T.™ compound (Sakura*, Finetek), frozen in liquid nitrogen and stored at -20°C before cryosectionning. Cryosections, 20 µm-thick, were obtained perpendicular to the gels, with a 2800N Jung Frigocut cryostat (Leika). The slides were air-dried at room temperature in a humidified chamber for at least 30 minutes and then blocked in PBS, 2% BSA, 0.05% Triton. Samples were incubated with anti-NGF monoclonal antibody (Promega, diluted 1/1000) overnight at 4°C. After extensive rinsing, the slides were incubated in the secondary antibody AlexaFluor® 594 (Invitrogen) and subsequently mounted with Prolong® Gold antifade reagent (Invitrogen). Stained collagen gel sections were observed on Zeiss Axioimager. To generate fluorescence intensity profiles, images were acquired on a DM 2500 Leica confocal microscope on three different sections. Four intensity profiles were generated with the ImageJ (National Institute of Health) freeware for each section at different places and distances were normalized by the total thickness of the gel section.

### 2.3 - Cell culture and differentiation

*PC-12 cells*. The rat pheochromocytoma cell line was obtained from the American Type Culture Collection (ATCC, Rockville, MD, USA). PC-12 cells were maintained in Dulbecco's Modified Eagle Medium (DMEM, Gibco®) supplemented with 2 mM glutamine, 5% heat-inactivated fetal bovine serum, 10% Horse serum and 1% antibiotics (penicillin, streptomycin), in a humidified incubator, under an atmosphere of 5% $CO_2$, at 37°C. For differentiation, PC-12 cells were seeded in DMEM without serum in 12-well plates at a density of 50,000 cells per well. Substrates for PC-12 cells differentiation were dense collagen matrices prepared as described above, or conventional collagen-coated coverslips as controls. Differentiation controls consist in adding 50 ng of NGF to the medium. NGF-releasing gels were prepared the day before seeding the cells, by impregnating the collagen scaffolds with 50 μl of a NGF solution at 10 μg/ml in DMEM overnight in a humid atmosphere. NGF-coupled collagen gels were used immediately after rinsing.

*SH-SY5Y cells*. SH-SY5Y neuroblastoma line was purchased from ATCC and routinely grown in DMEM-F12 with 2 mM glutamine, 10% fetal bovine serum, 1% antibiotics (penicillin, streptomycin), and fungizone® antimycotic (Gibco®, 2.5 μg/ml). SH-SY5Y cells were seeded on collagen substrates in 12-well plates at a density of 100,000 cells per well and maintained 24 hours in DMEM-F12 with 10% fetal bovine serum. SH-SY5Y cells were then differentiated in DMEM-F12 supplemented with 3% fetal bovine serum and 6.8 μM all-trans retinoic acid (Sigma). BDNF-releasing gels were prepared the day before seeding the cells, by impregnating the collagen scaffolds with 50 μl of a BDNF solution at 1 μg/ml in DMEM-F12.

BDNF-coupled collagen gels were used immediately after rinsing and compared to differentiation controls with 50 ng BDNF (Recombinant human BDNF, Gibco®).

**2.4 Cells immunofluorescent labelling and imaging**

Cells were fixed in paraformaldehyde 4% (Sigma) in PBS for 30 minutes after 4 days of differentiation for PC-12 cells and 6 days for SH-SY5Y cells. After fixation and washing in PBS, cells were blocked and permeabilized in 2% BSA and 0.1% triton X-100 in PBS for 20 minutes. Samples were incubated with primary antibody monoclonal anti-βIII-tubulin (mouse monoclonal, 1/200, Sigma) overnight at 4°C. After three washes in PBS, samples were incubated with the secondary antibody Cy3 Donkey anti-mouse IgG (1/400, Jackson ImmunoResearch Laboratories) for one hour at room temperature. During the last wash, nuclei were counterstained by incubating in DAPI (300 mM, Invitrogen) for 5 minutes before mounting with ProLong® Gold Antifade reagent (Invitrogen).
Fluorescence images were acquired on DM 2500 Leica confocal microscope. Images stacks were reconstructed with ImageJ.

**2.5 Neurite outgrowth measurement**

Neurites outgrowth was quantified on epifluorescence images (Zeiss Axioimager). Images of 5 fields per well were taken. Neurite growth was determined by manually tracing the length of (1) all the neurites which could be associated with a particular cell body and (2) the longest neurite per cell for all cells in a field that had an identifiable neurite and for which the entire neurite arbor could be visualized. We measured the length and total output (sum length) of only primary neurites. Neuronal processes were analyzed using the Axiovision LE 4.8 software. For each graph, neurite length data were generated from at least three independent experiments and more than 80 cells were counted for each experiment. Data were analyzed by Student's *t* test. We consider that a cell is differentiated when it possesses at least one neurite length superior or equal to two fold the cell body diameter.

## 3. Results and discussion

### 3.1 - Collagen biomimetic matrices

*- Grafting onto glass surfaces*

Our experience told us that relatively thick (50 µm) collagen hydrogels do not strongly adhere to glass and often detach from the coated coverslips during manipulation (gel preparation, culture, staining, mounting…), which can seriously impair cellular investigations. In what follows, we characterize the anchoring of dense fibrillar collagen matrices to glass surfaces, as schematically depicted in figure 1 through a combination of chemical coupling and physical gel formation. The first part of the procedure consists in covalently attaching collagen molecules to the glass surface. Contact angle measurements were performed after each step to determine the evolution of the surface energy as a first indicator of the grafting efficiency. Bare glass slides, *ie* without any modifications, have a mean water contact angle of 56±1°. After silanization with APTES and then coupling of glutaraldehyde, the contact angles were respectively 78±2° and 74±6°, indicating increased hydrophobicity attributed to the hydrocarbon chains of APTES and GA. After collagen immobilization, the water contact angle decreased back to 45±6°, consistent with the high hydrophilicity of most proteins. Grafting steps were followed at the molecular scale by PM-IRRAS. The silanized and glutaraldehyde-treated surfaces (Figures 2a and 2b) displayed a large positive band between 1460 and 1440 $cm^{-1}$ attributed to $CH_2$ bending, and a negative band at 1581 $cm^{-1}$ assigned to $NH_2$ bending. The negative sign of the last band indicates a specific orientation of the transition moment, perpendicular to the substrate surface. After addition of collagen and extensive washing (Figure 2c), the characteristic amide I (~1650 $cm^{-1}$) and amide II (~1550 $cm^{-1}$) bands of proteins are observed. Unlike for bulk collagen, where the amide I band is much stronger than the amide II [39], in the present case of coupled triple helices, both bands display nearly the same intensities. This unusual ratio indicates an anisotropic orientation of collagen molecules relative to the substrate. In the collagen triple helix structure [40], C=O and NH links are roughly perpendicular to the helix axis, resulting in respectively parallel and perpendicular amide II and amide I transitions moments. Considering surface selection rules of PM-IRRAS, the low amide I/amide II intensity ratio observed in our spectra can thus be attributed to a globally flat orientation of triple helices on the substrate in the dry state.

After a first monolayer of collagen was covalently attached to the glass surface, a thick gel was cast in a mould of defined thickness by pouring an acidic collagen solution at a chosen concentration (here 45 mg/mL) then submitted to ammonia vapors to induce fibrillogenesis [35, 41]. Ultimately, the gel grafting efficiency was assessed qualitatively by the sustained adhesion of the collagen matrices to the glass supports throughout the sample preparation steps and cell culture experiments. In an attempt at quantifying the functional gain of grafting the gels, we estimated the force necessary to detach them from the glass coverslip by measuring the tensile breaking strength for grafted samples compared to ungrafted ones. This was done by pulling on thick gels sandwiched between two coverslips, in the presence or absence of chemical grafting. In all cases, the rupture appeared to occur at the interface between the coverslip and the gel and not in the bulk of the gel. The breaking strength was shown to increase about six times, from approximately $5\pm0.2$ kN.m$^{-2}$ for ungrafted samples, to about $26\pm4.4$ kN.m$^{-2}$ for grafted ones. Ultimate tensile strength measured by other groups for similar collagen materials prepared *in vitro* being rather on the order of the MPa [42, 43], the interface appears weaker than the bulk fibrillar gel. This is probably because the density of physical reticulation nodes (native collagen fibrils) within the gel is much higher than the density of covalent bonds between the gel and the glass surface, resulting in a lower mechanical strength at the interface. In any case, the simple grafting process presented here proved sufficient to prevent detachment of the gels during the following experimental procedures.

*- Biomimetic structure and 3D organization*

Native collagen fibrils, with the typical 67 nm – periodic striation, are visible by SEM (Figure 3) on the surface of the dry collagen scaffolds. The fibrils have diameters *ca.* 100 nm and appear locally aligned. The samples thickness was estimated in SEM (edge view) under low vacuum conditions slightly below 40 µm after supercritical $CO_2$ drying. The difference with the height defined by the mould geometry (50 µm) can be attributed to the contraction of samples during fibrillogenesis and/or a partial collapse of the gel during supercritical $CO_2$ drying. However, there is no significant difference neither in structure nor in thickness, between gels supercritically dried in $CO_2$ immediately after chemical fixation, and gels previously freeze-dried and rehydrated (corresponding to the growth factor impregnation step) prior to supercritical drying. Overall, the gels appear dense but porous with a flat surface exposed to the cells to be seeded on top.

## 3.2 – PC-12 cells behavior on NGF-containing collagen hydrogels

Although type I collagen is a common substrate for the culture and neural differentiation of PC-12 cells [33, 44], we first assessed that they indeed survive and differentiate on these relatively soft hydrogels, at least compared to coated glass or plastic. When 50 ng NGF are added to the culture medium without serum, the proportion of cells showing the morphological features of differentiated cells after 4 days reaches about 70% (data not shown), close to the maximum value reported previously [44] with conventional coatings.

When NGF was included in the gel by simple impregnation, PC-12 cells also exhibited a neural phenotype with numerous elongated cellular processes and branching, as illustrated by β-3 tubulin staining in Figure 4A after four days of culture in serum-free medium. It is noteworthy that PC-12 cultured in the culture medium without NGF and in the absence of serum do not differentiate at all and most often die within a few days [45]. Therefore, as described previously for other hydrogels [46-48] or thin collagen coatings [49], active NGF is released from the thick matrix and stimulates neurite outgrowth. However, to reach a similar differentiation level (Figure 4B,C, *t-test* analysis), larger amounts (500 ng) of impregnated NGF were necessary compared to the condition where the growth factor is directly added to the medium (50 ng).

This difference can be in part ascribed to the release kinetic of NGF from the gel and its delivery to the cells. The quantification by ELISA of the time-dependant amount of native NGF released in the medium (Figure 5A) indicates that only ~1/3 (35%) of the initial impregnated amount is released and detectable by ELISA over the course of our experiments (4 days). This suggests that a significant proportion of the impregnated growth factor could be retained in the gel. Although to our knowledge, no specific interaction sequence has been identified, adsorption of NGF on collagen coatings has been reported before [50, 51]. The spontaneous deactivation of NGF, especially in dilute solutions [52] could also explain most of the difference between the loaded amount and the quantity measured in the medium after four days of release from the gel (personal data). Considering therefore that only part of the growth factor is accessible to the cells, and given the strong NGF dose-response dependence in the 0-50 ng range reported for PC-12 [45, 53, 54], the amount released with 50 ng of NGF impregnated in the gel (~10–15 ng) would be insufficient to obtain high differentiation levels. Therefore, to ensure optimum differentiation in all conditions, we chose to work in the present study, with an excess of immobilized growth factor (500 ng), which could probably be reduced by at least a factor of 3 and still induce a neural phenotype for a majority of PC-12 cells.

One major advantage of using thick dense collagen matrices to differentiate neural cells is that neurite outgrowth is in principle possible in 3D within the matrix supporting the cells. In these conditions we were able to observe, by confocal microscopy, differentiated cells with neurites growing several tens of microns into the depth of the gels (Figure 5B). Here, the growth factor necessary to the cells differentiation is delivered from the gel. It is absent from the medium when the cells are seeded, which results in a steep concentration difference between the interior and exterior of the gel. The slow release of NGF from the matrix should result in the formation of concentration gradients around the gel/medium interface [47], the bottom of the gel acting as a reservoir. However, such gradient is transitory and disappears once the system approaches equilibrium (plateau in the release kinetic), which occurs in about 24 hours. The chemotactic effect [8, 46] expected from the establishment of diffusion-induced NGF concentration gradients appears too limited [55] to guide massively the growth of neurites into the gel.

### 3.3 – Efficiency of growth factor covalently coupled to collagen matrices

We thus turned to the covalent coupling of the growth factor to enhance the chemotactic effect of the NGF included in the gel and foster 3D neuritic growth. The EDC/NHS [56, 57] coupling efficiency was evaluated by immuno-fluorescent staining of NGF on cryosections of collagen gels (Figure 6A-D). An exposure time of 1.4s is required to reveal the control gels without NGF (Figure 6A) or without linkers (Figure 6B), while NGF-coupled gels are overexposed (Figure 6C). The much larger fluorescence intensity observed for the former indicates efficient coupling, as confirmed by the colocalization of immuno-fluorescently labeled NGF with large isolated collagen fibrils observed in the differential interference contrast (DIC) mode (Figure 6F). Concentration profiles of collagen-coupled NGF were established by measuring the fluorescence intensity across the gel, *ie* along the direction perpendicular to the surface (Figure 6D-E). The density of coupled NGF increases moderately and gradually throughout the gel from the glass coverslip to the scaffold surface. Such gradient is not unexpected considering the simple one-way diffusion of NGF from the top of the gel to the bottom and its subsequent coupling to the NHS-activated collagen.

NGF-coupled collagen gels proved efficient to induce a neural phenotype in PC-12 cells. With 50 ng of NGF coupled to the gel (Figure 7A, 50 ng -c- gel) the differentiation rate is slightly increased with respect to control (Figure 7A, No linkers), but remains much lower (31.6%) than when the same amount is added in a soluble form to the culture medium (70.2%) (Figure 7A, 50 ng @ med)). Adding 500 ng of NGF to the matrix (Figure 7A, 500 ng -c- gel) restores a high level of differentiation (61.1%) with well developed neurites extending over several times the size of the cell body as illustrated in Figure 7B. The total length of neurites emitted by individual differentiated cells on 50 and 500 ng NGF-coupled collagen gels is also significantly increased compared to the control without linkers (Figure 7C). This clearly indicates that at least part of the bound growth factor is active. It should be noted that we did not prime the cells by exposing them to soluble NGF before seeding them on the matrices. Therefore, the neural phenotype was principally initiated by growth factor molecules coupled near the surface. As reported previously for NGF [55, 58], other growth factors or proteins [56, 59, 60], interaction of the active molecule covalently coupled to the matrix with corresponding membrane receptors is enough to induce cell survival and in some cases differentiation.

Yu *et al*. [58] evaluated that a 30 ng/cm$^2$ surface density of bound NGF yields a similar neural survival rate (~75%) as with 50 ng/ml soluble NGF, while a 10-fold reduction to 3 ng/cm$^2$ brought the survival rate down to less than 40%. In the present study, the growth factor is coupled to the whole gel and only the surface is initially accessible to cells. Given the rugosity of the dense collagen matrix surface and considering, for the sake of rough estimation, that about one tenth (5 μm-deep) of NGF is directly accessible yields NGF surface densities of about 2 ng/cm$^2$ (50 ng coupled) and 20 ng/cm$^2$ (500 ng coupled). Although the NGF dose-response of PC-12 survival and differentiation rates are probably different, this dependence might explain the slightly lower differentiation rate observed with 500 ng coupled NGF as compared to its addition at 50 ng/ml in the medium.

We also observed neurites growing in the depth of the gel (Figure 7D). Although this might not be considered strictly as "neurite guidance" [8, 55, 61, 62], the topological and biochemical cues of native collagen fibrils combined to the presence of coupled growth factors are therefore enough for neurites to grow within the gel.

### 3.4 – Neuroblastoma cells culture and differentiation

To demonstrate that our cell differentiation support is amenable to a variety of neural cells, we investigated the behavior of human neuroblastoma cells (SH-SY5Y). This cell line, derived from tumors of the human peripheral neural system, was chosen because it presents many of the characteristics of primary cultures of neurons and was often used to study molecular and cellular features of neurodegenerative diseases like AD [63, 64]. When seeded on collagen hydrogels in the presence of 6.8 µM retinoic acid (RA), about 30% of the SH-SY5Y cells are differentiated, even in the absence of BDNF (Figure 8A and B, 0 ng BDNF). This proportion rises to more than 60% upon addition of 50 ng of BDNF in the medium (Figure 8A and B, 50 ng @ med). The same amount impregnated in the collagen matrix (Figure 8A and B, 50 ng @ gel) also induces a significative increase of differentiated cells compared to the control condition treated with RA only. Neurite length is slightly increased when BDNF is added either in the medium or in the gel by impregnation (Figure 8C). Surprisingly, the BDNF release profile determined by ELISA (Figure 8D) shows a maximum at 24 hours (100 ng; 20% of the loaded amount), followed by a slow decrease. We determined that the proportion of native (ELISA-detectable) BDNF decreases exponentially with time with an exponent value ($\sim e^{-kt}$, $k = 0.04$ day$^{-1}$) very close to that found previously by Callahan *et al.* (0.03 day$^{-1}$) [65]. It also appears that BDNF largely adsorbed onto the polystyrene plastic walls of the culture plate despite the presence of BSA-containing serum in the release medium. This strong adsorption, which does not occur in polypropylene microtubes (data not shown), amounts to about 300 ng, *ie* ~60% of the initial amount in solution after 24 hours. The combination of spontaneous BDNF degradation and its adsorption to the plate well surface explain the non-monotonic release profile in figure 8D.

BDNF chemically coupled (50 ng) to the gels is also able to stimulate the differentiation of SH-SY5Y as illustrated in figure 8A. The differentiation rate (Figure 8B) and neurite length (Figure 8C) are slightly increased compared to the control with RA only and are similar to those obtained with the same amount of soluble growth factor. Treatment of SH-SY5Y with RA at concentrations in the µM range is known to potentiate their responsiveness to the subsequent addition of BDNF [66-69]. It is therefore expected that, provided enough growth factor is available for the cells, typically 50-100 ng/ml [66-68], it induces the production of numerous long processes indicative of neural differentiation. These data again indicate that the EDC/NHS covalent coupling method enables the interaction of the growth factor with its membrane receptors and activation of the appropriate cellular signalization pathways. Interestingly, in all experiments we mostly observed N-type cells and very few S-type cells, although the cells were kept in the presence of RA and serum (3% Horse Serum) during their treatment with BDNF [67, 70].

**Conclusions**

We functionalized calibrated dense fibrillar collagen matrices to differentiate cell types widely used as models to study neural behavior. We successfully cultured and induced PC-12 and SH-SY5Y cells into a neural phenotype on thick (*ca.* 50 μm) collagen hydrogels with appropriate growth factors, respectively NGF and BDNF. When impregnated in dense matrices and delivered to the cells by simple diffusion-limited release, the trophic effect of both neurotrophins was enough to reach high proportions of differentiated cells (about 45 to 65 %). In addition, both NGF and BDNF chemically conjugated to the collagen matrices were also able to induce a neural phenotype, suggesting that intracellular signaling was efficiently triggered by their interaction with membrane receptors. For both impregnated and conjugated GF, we observed neurites growing several tens of microns into the depth of the fibrillar gels.

The scaffolds described herein were prepared by simple self-assembly processes with readily adaptable structure (porosity, collagen density, fibril size…) and controlled geometry. They are particularly suitable for the study of neurite growth in a biomimetic reconstructed matrix and direct observation by confocal microscopy. The simple and versatile coupling chemistry and preparation protocols make it easy to elaborate 3D cell microenvironments adapted for cellular investigations, especially those requiring vertical compartmentation. For instance, it should be a valuable tool to study the relationship between Alzheimer's diseases and neurite dystrophy, which has been shown to be at least two-prone. First, dystrophic neurites are associated to degenerating neurons and to amyloid β-peptide (Aβ) aggregates in senile plaques [71-73]. Second, it was shown in previous works using compartmented systems that neurites isolated from the cell body treated with Aβ aggregates undergo degeneration followed by cell death [5]. However, investigating the specific role of such neurite / Aβ interactions in neuron death with 2D culture systems is hampered by the difficulty to truly treat the neurites separately from the cell body, while permitting their growth from differentiating cells. To test this hypothesis, we are currently investigating the adverse effects of Aβ-amyloid aggregates included in a 3D biomimetic collagen matrix, in the vicinity of growing neurites.

**Acknowledgments**

The authors are grateful to Daniel Brunel for fruitful discussions and advice regarding the coupling chemistry and to Bernard Desbat for the PM-IRRAS measurements. We wish to thank Vicky Diakou for her assistance at the Montpellier RIO Imaging facility.

# Figure Captions

**Figure 1**: Collagen grafting and thick dense matrices elaboration. **(1)** Conventional glass coverslips were functionalized by silanization with APTES and glutaraldehyde before covalent grafting of collagen triple helices. **(2)** An acidic concentrated collagen solution was cast on the top of the grafted coverslip to form, after **(3)** exposition to $NH_3$ vapors, a 50 µm-thick fibrillar collagen gel (4).

**Figure 2**: PM-IRRAS spectra of the glass coverslip surface after **(a)** silanization with APTES, **(b)** glutaraldehyde reaction and, **(c)** collagen triples helix coupling.

**Figure 3**: Representative SEM images of fibrillar collagen scaffolds after chemical fixation and supercritical drying. **A** and **B**: Samples kept hydrated after fibrillogenesis under ammonia vapors. **C** and **D**: Samples lyophilized after gel formation and then impregnated with phosphate buffer. **A** and **C**: Surface pictures, scale bar 1 µm. **B** and **D**: Edge view visualized under low vacuum, scale bar 40 µm.

**Figure 4**: Effect of NGF added to the medium or impregnated within gels on PC-12 cells differentiation without serum. 50 ng of NGF were added to the medium (50 ng @ med) or 500 ng included in the gel by impregnation (500 ng @ gel). **A**: Immunofluorescence analysis of differentiated PC-12 cells after β-3 tubulin (red) and DAPI (blue) staining. Images were acquired in confocal microscopy (Z-projections of 4 confocal optical sections, interval 0.5 µm). Scale bar 20 µm. **B**: Quantitative analysis of undifferentiated and differentiated PC-12 cells. **C**: Total neurite length per cell was measured and expressed as mean ± SEM.

**Figure 5:** **A**: Release profile of NGF from fibrillar collagen gels measured by ELISA. The quantity (ng) and proportion of the initial amount expressed as mean ± SD were obtained with three distinct experiments with sampling performed in triplicate. **B**: Immunofluorescence images of differentiated PC-12 cells after β-3 tubulin (red) and DAPI (blue) staining. Images were acquired in confocal microscopy (49 confocal optical sections, interval 0.5 µm). Top, Z-projection of confocal slices; bottom, side-view reconstruction of the 49 confocal sections showing neurite growth through the gel. Scale bar 20 µm.

**Figure 6**: Spatial distribution and efficiency of NGF coupling to collagen gels. **A-D:** Immunofluorescence images of cryosectionned dense collagen matrices stained with an anti-NGF antibody: control gel without NGF (**A**), control gel without crosslinkers (**B**) and NGF-coupled collagen gel (**C-D**). The images were obtained with the same laser excitation and an exposure time of 1.4 s (A-C) or 0.3 s (D). Scale bars 50 μm. **E**: Fluorescent intensity profile of NGF coupled to collagen as a function of the normalized distance from the coverslip surface (Grey curves: ± 1σ). **F**: Co-localization of the anti-NGF immunofluorescent labeling and collagen fibers observed in the DIC mode. Scale bar 5 μm.

**Figure 7**: Effect of collagen coupled NGF on PC-12 cells differentiation. **A**: Quantitative analysis of differentiated and undifferentiated PC-12 cells. **B**: Immunofluorescence images of differentiated PC-12 cells after β-tubulin (red) and DAPI (blue) staining. Images were visualized by confocal microscopy (Z-projections of 4 confocal optical sections, interval 0.5 μm). Scale bar 20 μm. **C**: The total neurite length per cell was measured and expressed as mean ± SEM. Asterisks indicate significant differences from control without crosslinkers following student *t-test* analysis. (* $p<0.05$, ** $p<0.01$ *** $p<0.001$). **D**: Immunofluorescence reconstructed (49 confocal optical sections, interval 0.5 μm) side-view image of differentiated PC-12 cells after β-tubulin staining. Scale bar 20 μm.

**Figure 8**: Effect of BDNF in the medium, impregnated or covalently coupled into the gels on SH-SY5Y cells differentiation. SH-SY5Y cells were cultured with 6.8 μM retinoic acid. BDNF (50 ng) was added to the medium (@ med), included in the gel by impregnation (@ gel) or chemically conjugated (-c- gel). **A**: Immunofluorescence images of differentiated SH-SY5Y cells after β-3 tubulin (red) and DAPI (blue) staining. Images were visualized by confocal microscopy (Z-projections of 5 confocal optical sections, interval 0.5 μm). Scale bar 50 μm. **B**: Quantitative analysis of differentiated and undifferentiated SH-SY5Y cells. Asterisks indicate significant differences from control without BDNF following student *t-test* analysis (* $p<0.05$). **C**: The total neurite length per cell was measured and expressed as mean ± SEM. **D**: Release profile of BDNF from fibrillar collagen gels measured by ELISA. The quantity (ng) and the proportion of the initial amount released into the medium (mean±SD) were obtained with three distinct experiments and sampling performed in triplicate.

# FIGURES

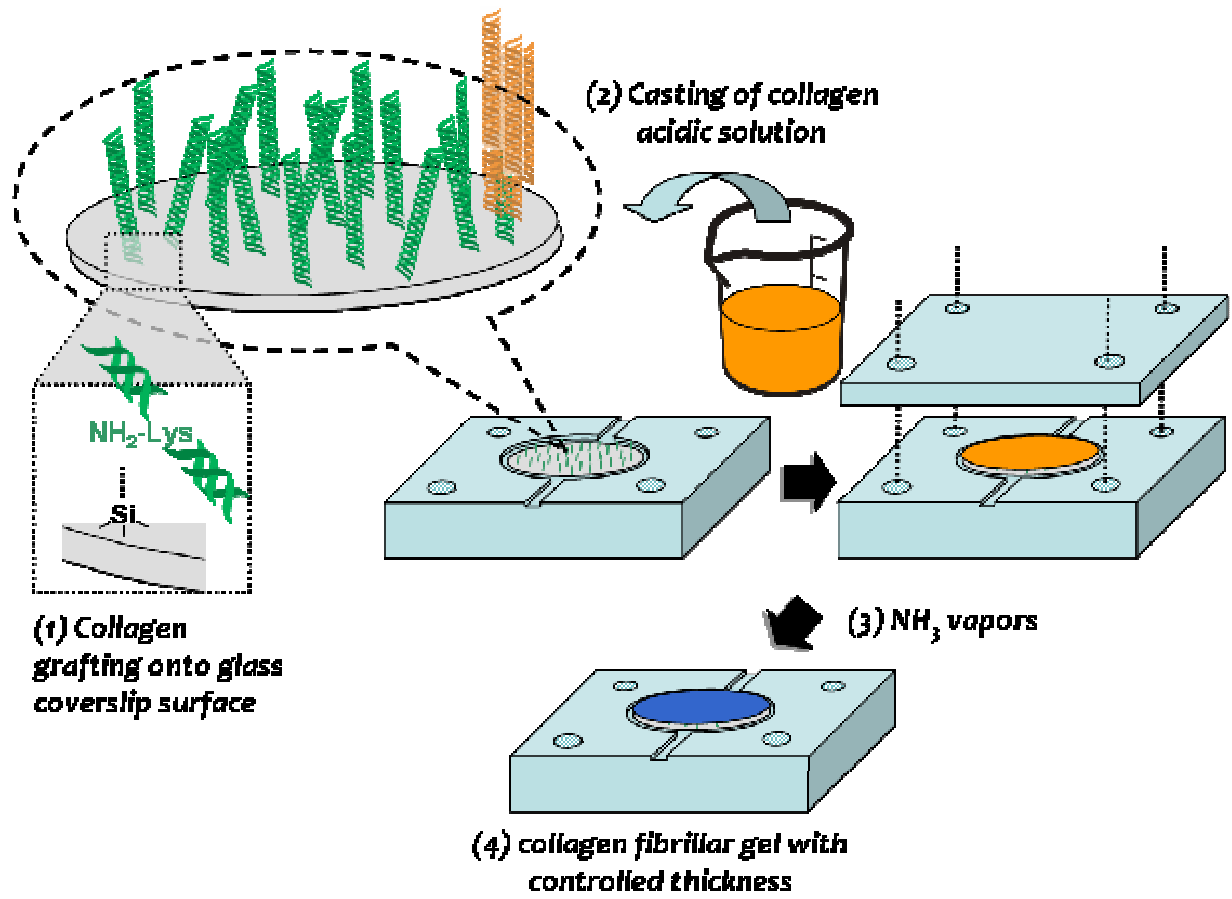

**Figure 1**

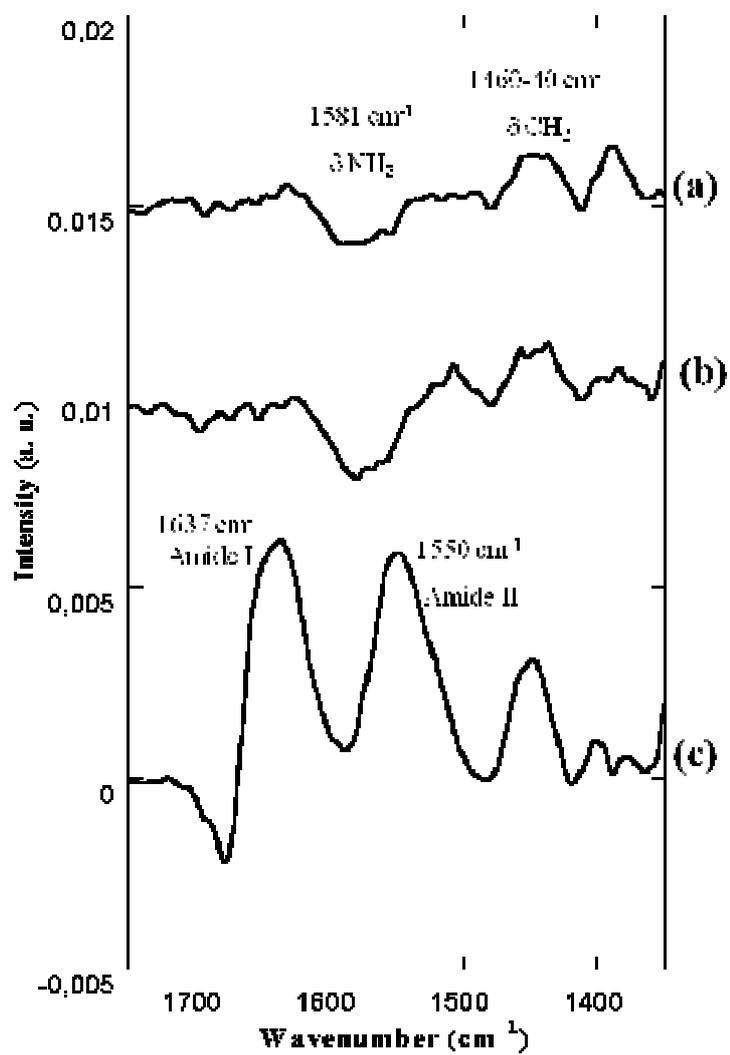

**Figure 2**

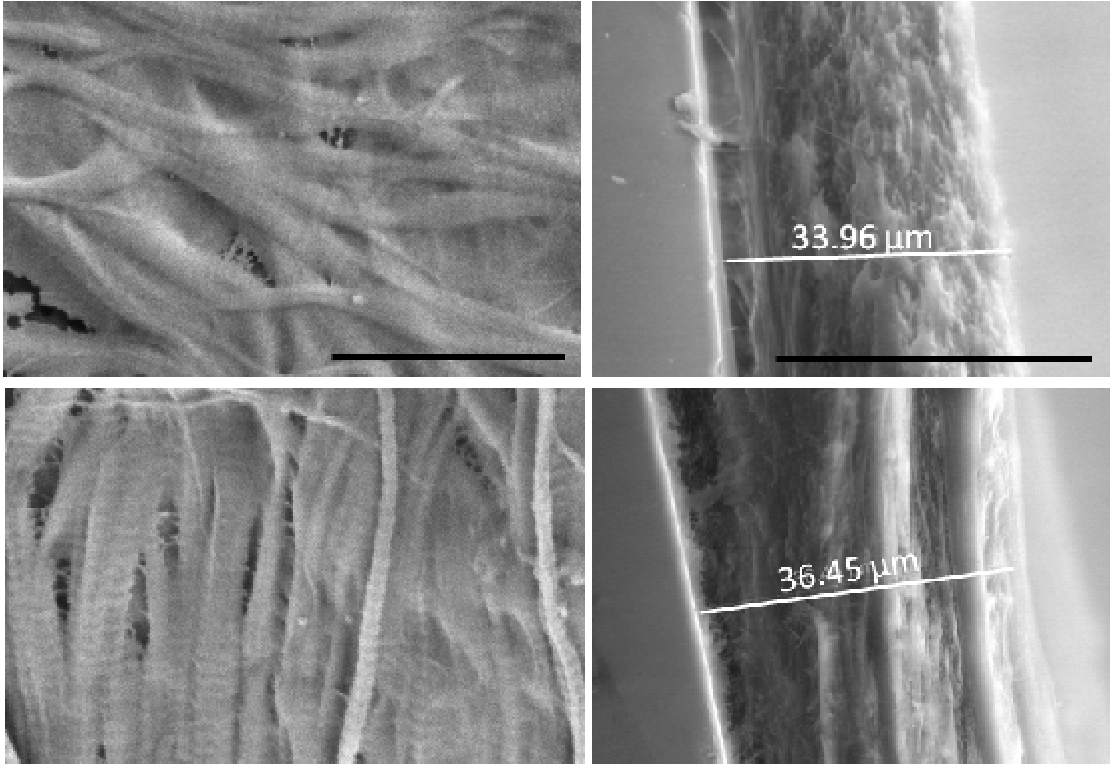

**Figure 3**

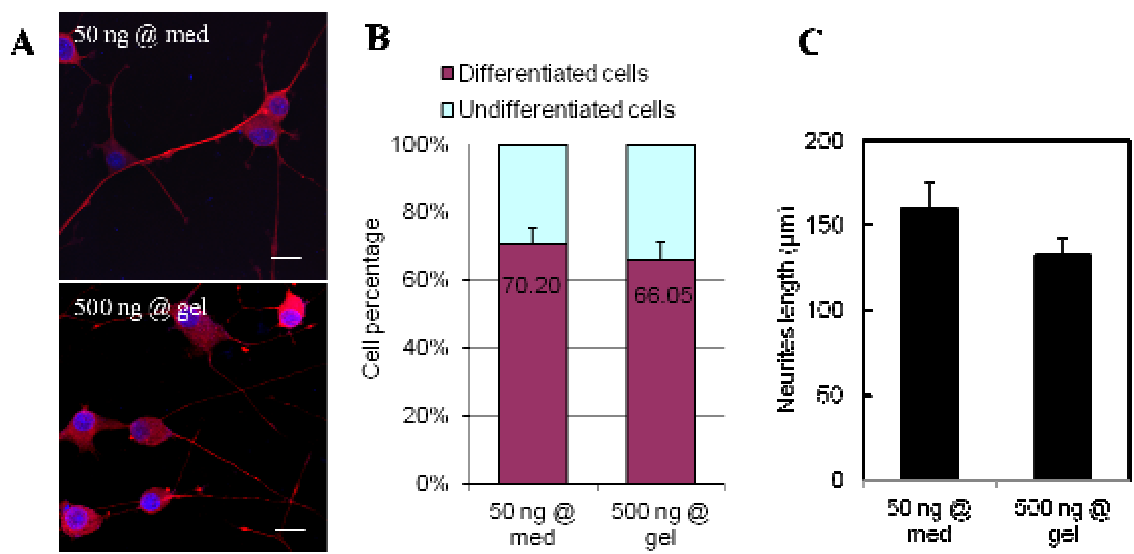

**Figure 4**

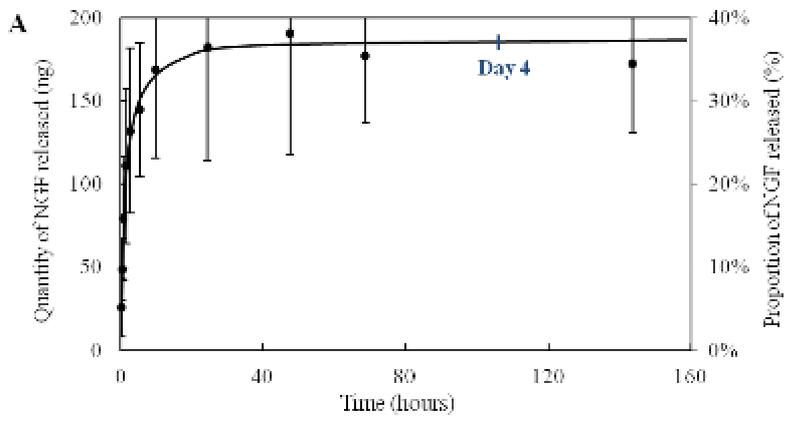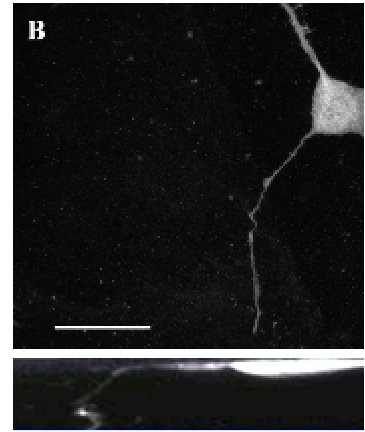

**Figure 5**

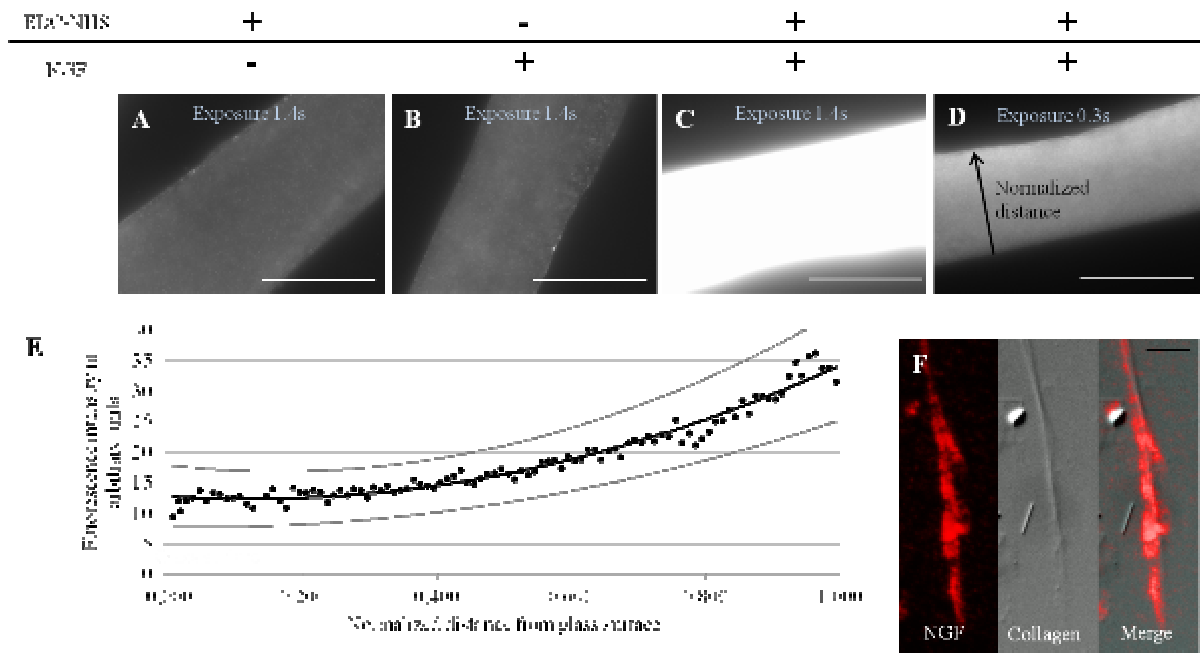

**Figure 6**

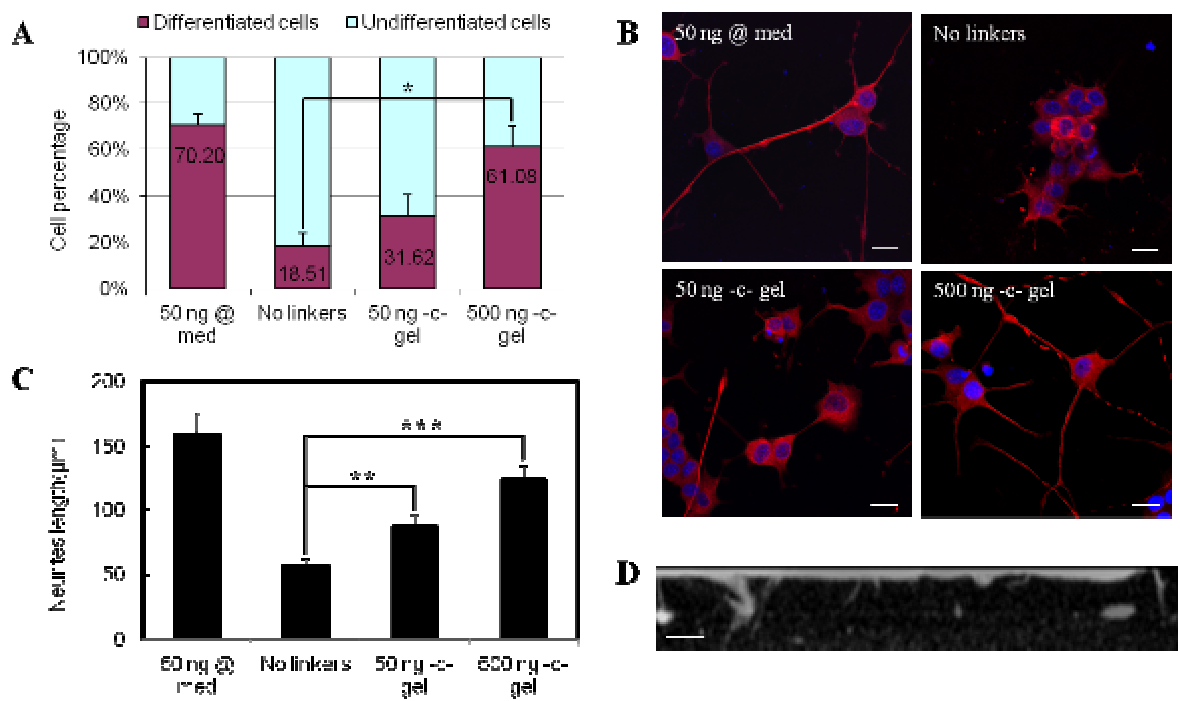

**Figure 7**

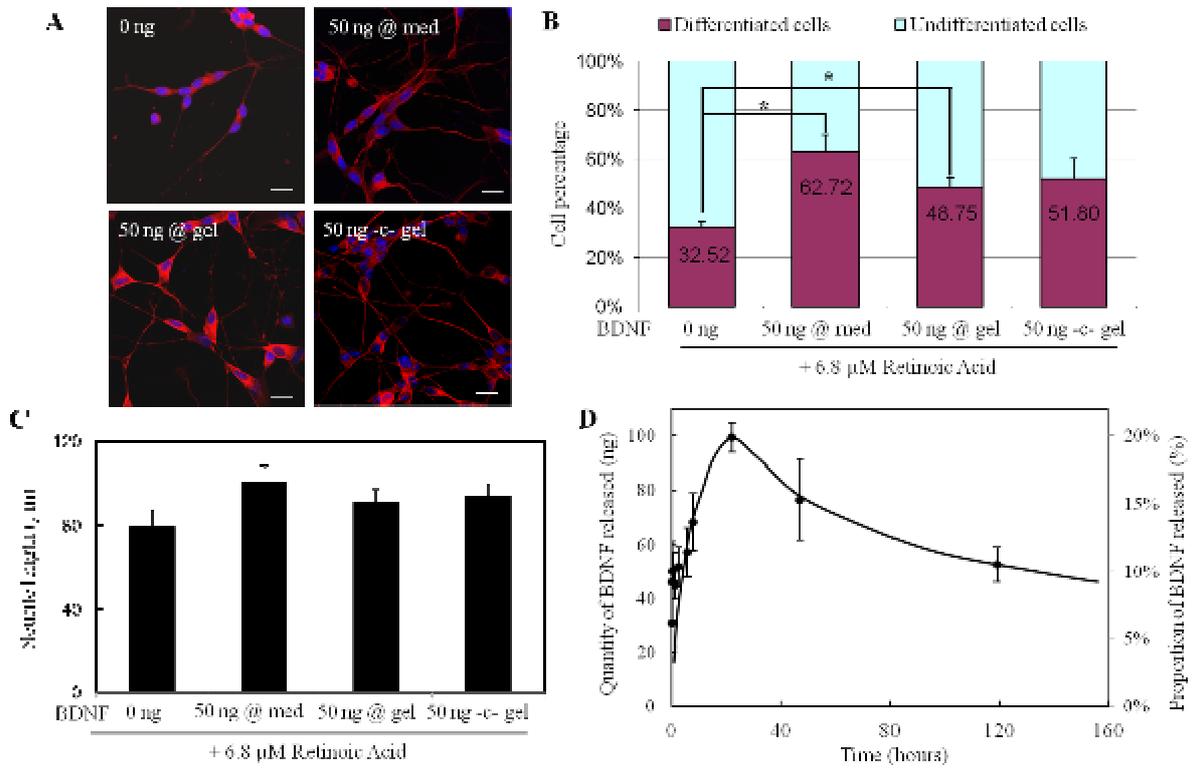

Figure 8